\documentclass{epl}

\usepackage{color}
\usepackage{amssymb}
\usepackage{graphics}

\title{Response maxima in time-modulated turbulence: \\Direct Numerical Simulations}
\shorttitle{Response maxima in time-modulated turbulence}

\author{Arkadiusz K. Kuczaj\inst{1}\thanks{e-mail: \email{a.k.kuczaj@utwente.nl}} \and Bernard J. Geurts\inst{1,2} \and Detlef Lohse\inst{3}\thanks{e-mail: \email{d.lohse@utwente.nl}}}
\shortauthor{Kuczaj, Geurts, and Lohse}

\institute{
  \inst{1} {Department of~Applied Mathematics - University of~Twente, \\P.O. Box 217, 7500 AE Enschede, the Netherlands} \\
  \inst{2} {Fluid Dynamics Laboratory - Eindhoven University of~Technology, \\P.O. Box 513, 5300 MB Eindhoven, the Netherlands} \\
  \inst{3} {Department of Applied Physics - University of~Twente, \\P.O. Box 217, 7500 AE Enschede, the Netherlands}
}

\pacs{47.27.Rc}{Turbulence control} \pacs{47.27.Eq}{Turbulence simulation and modeling} \pacs{47.27.Gs}{Isotropic turbulence; homogeneous turbulence}

\begin{document}

\maketitle

\begin{abstract}
The response of turbulent flow to time-modulated forcing is studied
by direct numerical simulations of the Navier-Stokes equations.
The~large-scale forcing is modulated via periodic energy input
variations at frequency $\omega$. The response is maximal for
frequencies in the~range of the~inverse of the~large eddy turnover
time, confirming the mean-field predictions of von der Heydt,
Grossmann and Lohse (Phys. Rev. E 67, 046308 (2003)). In accordance
with the~theory the~response maximum shows only a small dependence
on the~Reynolds number and~is also quite insensitive to
the~particular flow-quantity that is monitored, e.g., kinetic
energy, dissipation-rate, or Taylor-Reynolds number. At sufficiently
high frequencies the~amplitude of~the~kinetic energy response
decreases as $1/\omega$.
For frequencies beyond the
range of maximal response, a~significant change in phase-shift
relative to the time-modulated forcing is observed.
\end{abstract}

\section{Introduction}

Recently, response maxima in time modulated turbulence have been
predicted within a mean field theory of turbulence \cite{hey03a}.
Subsequently, such response maxima were found \cite{hey03b} in
numerical simulations of simplified dynamical turbulence models such
as the~GOY model \cite{bif03,bohr,kad95} or the reduced wave vector
approximation (REWA) \cite{egg91a,gnlo92b,gnlo94a,gnlo94b}. However,
these response maxima computed in \cite{hey03b} were not pronounced
at all, due to the approximate treatment of the small scales in
either of these approaches. Indications of response maxima resulting
from time-modulated forcing have subsequently also been seen in
experiment \cite{cad03}. The~experimental observations were done by
introducing a time-dependent swirl to fluid in a closed container
and monitoring the energy-dissipation-rate. The selected set-up did
not allow to identify possible flow-structuring under resonance
conditions, nor to conclusively distinguish such resonance phenomena
from flow-organization associated with the~size of~the~container.

The purpose of this paper is to complement these theoretical,
numerical, and experimental observations by direct numerical
simulations (DNS) of turbulence, subject to time-modulated
large-scale forcing. In a turbulent flow whose large-scale forcing
is periodically modulated in time, all typical flow-properties
develop a complex time-dependence. However, averaging such turbulent
time-dependence, conditioned on the phase of the periodic
modulation, yields a clear and much simpler periodic pattern
\cite{hey03b}. The dependence of the conditionally averaged response
on the frequency of the modulation may be quantified by monitoring
changes in flow-properties such as total energy, dissipation-rate,
or Taylor-Reynolds number. In case of~a~fast modulation with
a~frequency $\omega \gg \omega_L$, where $\omega_L$ is the inverse
large eddy turnover time, only a modest effect on the flow is
expected, or none at all. Likewise, if $\omega \ll \omega_L$ the
modulation is quasi-stationary and the flow may be expected to
closely resemble the~corresponding unmodulated case. In between
these extremes a more pronounced response may develop, which is
the~subject of this investigation.

The DNS approach allows to investigate in detail the response of
turbulent flow-properties to periodic modulation of the forcing. In
particular, we present an extensive parameter-study involving a
large range of modulation frequencies for two different Reynolds
numbers, and~establish response maxima in a variety of
flow-properties. The response is found to be significantly increased
at modulation frequencies on the order of the inverse of the
eddy-turnover time. Near resonance, the `activity' of the turbulent
flow is found to be considerably higher than in the unmodulated
case.
At high frequencies $\omega$ the~amplitude of
the~modulation-specific response of the kinetic energy is found to
uniformly decrease to zero as $\omega^{-1}$. This type of external
control of turbulence may offer new opportunities with relevance to
technological applications.

The organization of this paper is as follows. We first introduce the
computational flow-model in more detail. Subsequently, an overview
of the ensemble averaging procedure and~data extraction is given.
Then the main result,
the response of various flow properties to time-modulated forcing,
 is presented.
The paper ends with a summary and conclusions.

\section{Computational flow-model}
The full Navier-Stokes equations for incompressible flow are
numerically solved in a periodic flow-domain with a pseudo-spectral
code. In spectral space, the~Navier-Stokes equations read
\begin{equation}\label{eq:ns}
\left[ {\frac{\partial }{{\partial t}} + \nu |\mathbf{k}|^2 }
\right]u_\alpha ({\bf{k}},t) = M_{\alpha \beta \gamma
}({\bf{k}})\sum\limits_{{\bf{p}} + {\bf{q}} = {\bf{k}}}
{u_\beta({\bf{p}},t)u_\gamma  ({\bf{q}},t)} + F_\alpha ({\bf{k}},t),
\end{equation}
with $M_{\alpha \beta \gamma } ({\bf{k}}) = \frac{1}{2\imath}
\Big({k_\beta D_{\alpha \gamma } ({\bf{k}}) + k_\gamma  D_{\alpha
\beta } ({\bf{k}})} \Big)$, with $D_{\alpha \beta }({\bf{k}})  =
\delta _{\alpha \beta } - {k_\alpha k_\beta }/{|\mathbf{k}|^2 }$.
Here, $\nu$ is the~kinematic viscosity, $u_\alpha ({\bf{k}},t)$ is
the Fourier-coefficient of the velocity field at wave vector
$\bf{k}$ and time $t$ and $F_\alpha$ is the time-modulated forcing.

First, we recall that traditional agitation of the large-scale
structures in a turbulent flow may be achieved by introducing a
forcing term restricted to wave vectors with {\mbox{$|{\bf{k}}| \leq
k_F$}}, i.e., identifying a forcing-range through the upper-limit
$k_F$. Specifically, we force the turbulence similarly as in
\cite{gnlo92b,gho95},
\begin{equation}\label{eq:f}
 f_\alpha  ({\mathbf{k}},t) = \frac{{\varepsilon _w }} {N_F}\frac{{u_\alpha  ({\mathbf{k}},t)}} {{\left| {{\mathbf{u}}({\mathbf{k}},t)} \right|^2 }}~~~~~;~~~~
 |{\bf{k}}|<k_F
\end{equation}
where $\varepsilon_{w}$ is the constant energy injection rate and
$N_F= N_F(k_F)$ is the total number of forced modes.
For convenience, the wave vectors are grouped in spherical shells
with the $n$-th shell containing all modes such that $(n - 1/2) <
|{\bf{k}}| \leq (n + 1/2)$. We applied large-scale forcing either in
the first shell at $n=1$ (i.e., $k_F=3/2$ which implies $N_F=18$,
the case considered in \cite{hey03b}) or in the first two shells
(i.e., $k_F=5/2$ which implies $N_F=80$). The second step in
specifying the forcing $F_{\alpha}$ introduces the periodic time
modulation
\begin{equation}\label{eq:ft}
F_\alpha ({\mathbf{k}},t) = f_\alpha ({\mathbf{k}},t)\Big( 1 + A_F
\sin ( \omega t)  \Big),
\end{equation}
where $A_F$ is the amplitude of modulation and $\omega$ its angular
frequency. The modulated forcing  corresponds to a total energy
input rate which oscillates around $\varepsilon_w$ with amplitude
$A_F$,
\begin{equation}
{T}_{F}(\omega,t) = \sum_{\mathbf{k}}
{u_\alpha^*({\bf{k}},t)F_\alpha ({\bf{k}},t)} =\varepsilon _w \Big(
1 + A_F \sin ( \omega t)\Big). \label{eq:et}
\end{equation}
The lengths and time scales of the numerical simulation are chosen
by picking $L=1$ for the~box-size in physical space and
$\varepsilon_w = 0.15$ for the energy injection rate. The Reynolds
number is then determined by the dimensionless viscosity $\nu$.
Choosing $\nu^{-1} =1060.7$ and $\nu^{-1} = 4242.6$ result in
respective approximate Taylor-Reynolds numbers $R_{\lambda} \cong
50$ and $R_{\lambda} \cong 100$. We use
these two cases  as
references denoted by $R_{50}$ and $R_{100}$. The spatial resolution
needed may be estimated by requiring $k_{max}\eta > 1$ \cite{esw88}
with $\eta$ the~Kolmogorov dissipation scale and $k_{max}$
the~highest wavenumber included in the spatial discretization.
For~$R_{50}$ case a~resolution of~at~least $N^3 = 64^{3}$
computational points is required while for~$R_{100}$ a~higher
resolution of $192^{3}$ points is necessary. The~latter poses
a~strong computational challenge in view of the extensive ensemble
averaging and large number of modulation frequencies. However, it
was found that many large scale quantities, such as the total
energy, do not depend too sensitively on resolution. As an example,
a~resolution of $64^{3}$ points corresponds to $k_{max}\eta \approx
0.4$ for the
$R_{100}$ case. Still, this resolution is quite adequate
for studying the response of total energy. This was verified by
repeating the analysis at a selection of modulation frequencies with
resolutions $128^{3}$ and $192^{3}$. The predictions of~quantities
that rely more on small scales, such as the dissipation-rate,
contain a higher numerical uncertainty for $R_{100}$ case and
$64^{3}$ computational points, but~still allow a~clear
interpretation of the main turbulence response. This was separately
assessed using the~higher resolution data at selected characteristic
frequencies.

The direct numerical simulation for the unmodulated case starts from
an initial condition that was generated on the basis of the Pao
spectrum \cite{pop00}. We adopt exactly the same initial conditions
as in \cite{mey03} which allow a separate validation of the
simulations. Explicit second order compact-storage Runge-Kutta
time-stepping \cite{geu03} with fully de-aliased pseudo-spectral
discretization is used. The~unmodulated turbulent flows provide the
point of reference for investigating the effect of modulated
forcing, to which we turn next.

\section{Averaging procedure and simulation setting}

\begin{figure}[hbt]
\begin{center}
\includegraphics[width=0.45\textwidth]{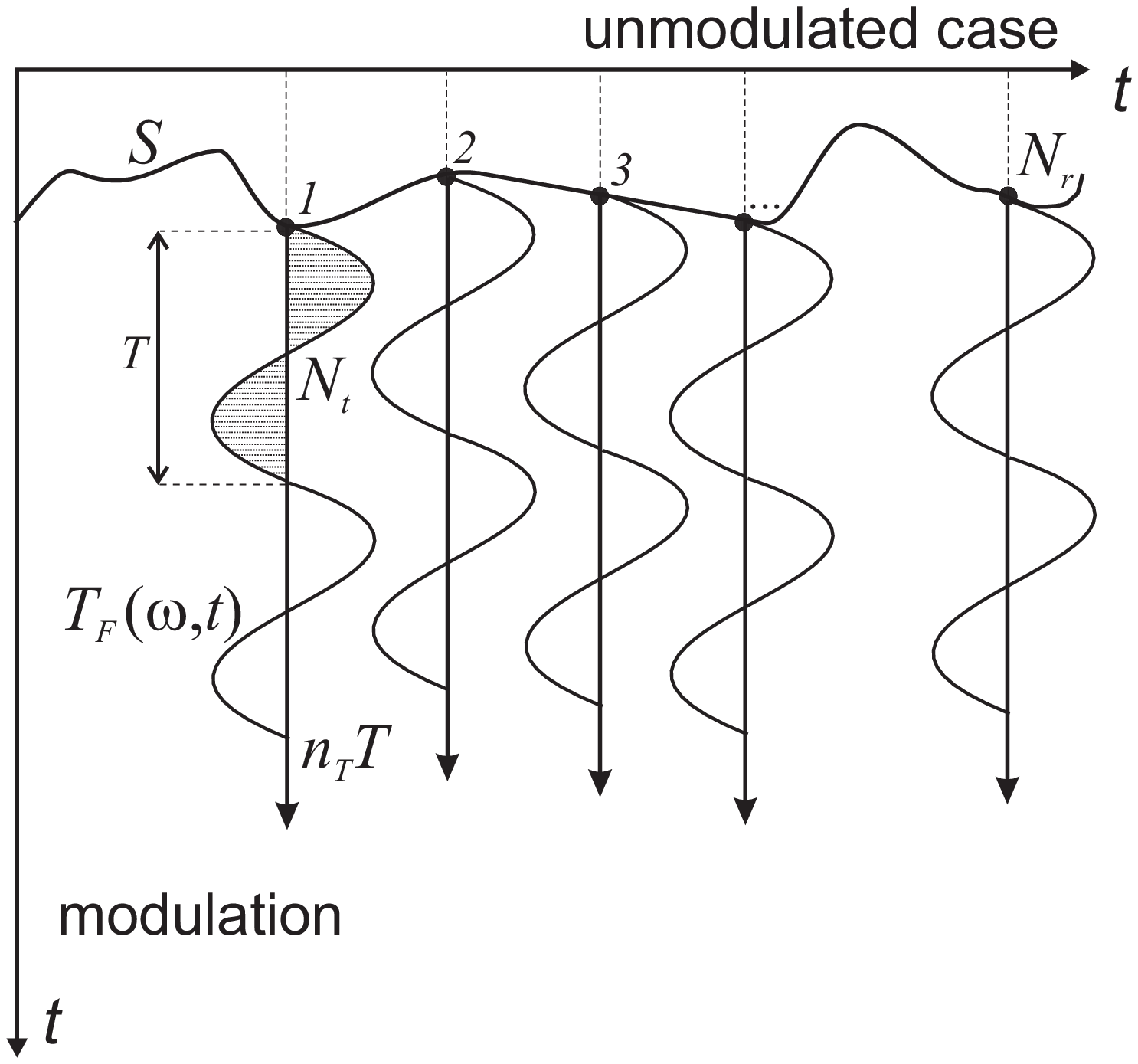}(a)
\includegraphics[width=0.45\textwidth]{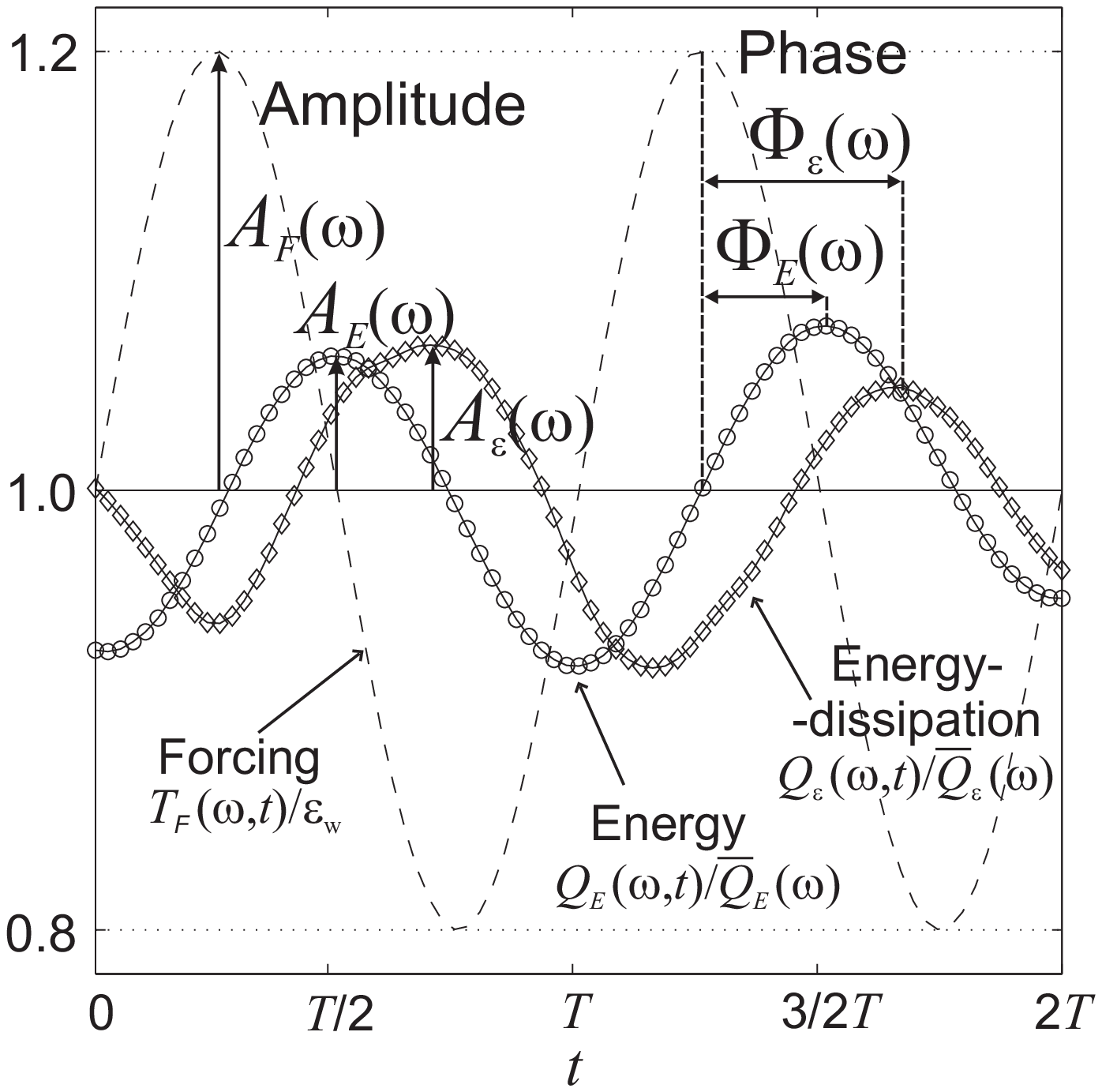}(b)
\end{center}
\caption{Procedure of data generation to compute the conditional
average (a) and the amplitudes $A_F(\omega)$, $A_E(\omega)$,
$A_\varepsilon(\omega)$ and phase-shifts $\Phi_F(\omega)\equiv 0$,
$\Phi_E(\omega)$, $\Phi_\varepsilon(\omega)$ of the forcing
$T_F(\omega,t)$ (dashed line), the energy $Q_E(\omega,t)$
(labeled~$\circ$), and the energy-dissipation-rate
$Q_\varepsilon(\omega,t)$ (labeled~$\diamond$) normalized by their
respective means $\overline{T}_F=\varepsilon_w$,
$\overline{Q}_E(\omega)$, and $\overline{Q}_\varepsilon(\omega)$
obtained from simulations at the modulation frequency $\omega=0.8
\pi$ (b).}

\label{fig:timemod0}

\end{figure}

In order to analyze the response to a~time-modulated forcing, the
precise extraction of the amplitude and phase of the conditionally
averaged variations is a key issue. Two steps can be distinguished,
i.e., the computation of~the~conditionally averaged signal
itself and the subsequent determination of amplitude and phase
characteristics of this signal, see Figure~\ref{fig:timemod0}. These
steps are discussed and illustrated next.

We adopt ensemble averaging to determine the conditionally averaged
signal $S(\omega,t)$, where $S(\omega , t)$ is the total energy $E(\omega,t)$,
the Taylor-Reynolds number $R_\lambda(\omega,t)$
or the energy
dissipation rate $\varepsilon(\omega,t)$.
Ensemble averaging
 requires a sufficiently large sample of
statistically independent signals $\{ S_{j}(\omega,t) \}$ to be
generated. The computational approach is summarized in
Figure~\ref{fig:timemod0}(a) and involves two main steps. Firstly,
we compute the unmodulated flow and store $N_r$ realizations of the
turbulent solution corresponding to $t > 10$. The latter condition
allows transients related to the initial condition to become
negligible. The time-separation between these snapshots is larger
than two eddy-turnover times. Subsequently, each of these $N_{r}$
realizations was taken as the initial condition for a simulation
with time-modulated forcing at a particular frequency $\omega$. This
provides $N_{r}$ sample signals which need to be averaged to obtain
the conditionally averaged signal $S(\omega,t)$. Repeating this
procedure for a range of frequencies yields the total response
characteristics. Given the conditionally averaged response signal
$S(\omega,t)$, there are various ways in which amplitude and phase
information can be extracted. In \cite{cad03} the signal
$S(\omega,t)$ was first averaged over time to yield
${\overline{S}}(\omega)$. Subsequently, the normalized variation
defined as $Q^{(a)}_S(\omega,t)=S(\omega,t)/{\overline{S}}(\omega)$
was studied using the Fourier-transform~($\mathcal{F}$) in which
time $t$ is transformed into frequency $f$. Correspondingly, the
power amplitude spectrum $\widehat
Q^{(a)}_S(\omega,f)=\mathcal{F}\big(Q^{(a)}_S(\omega,t)-1\big)$ can
be obtained which assumes a maximum value $A_{S}(\omega) =
\max\{|\widehat Q^{(a)}_S(\omega,f)|\}|_{f=f_{A}(\omega)}$, as
denoted in Figure~\ref{fig:timemod0}(b) for forcing $A_F(\omega)$,
total energy $A_E(\omega)$, and energy-dissipation-rate
$A_{\varepsilon}(\omega)$. The~maximum ${A}_{S}(\omega)$ as the
amplitude at dominant frequency can be used to quantify the response
as function of the modulation frequency~$\omega$. This~approach is
accurate if Fourier-transform is applied to an integer number of
modulation periods. The~method used in \cite{hey03b} is based on a
fitting procedure in which it is assumed that $S(\omega,t) \approx
\overline{S}+A_S\sin\big(\omega t+\Phi_S\big)$. The dependence of
the parameters $\{ \overline{S}, A_S, \Phi_S\}$ on $\omega$ may be
obtained from a least squares procedure. This evaluation method
assumes that the conditionally averaged signal has the same
frequency as the forcing.

At modest ensemble-size $N_{r}$ it is beneficial to explicitly
incorporate variations in the unmodulated reference signal to
improve the data-evaluation. This motivates an alternative method in
which we determine $N_{r}$ sample signals $\{ S_{j}(\omega,t) \}$
corresponding to the modulated case, as well as $N_{r}$ unmodulated
signals $\{s_{j}(t)\}$ that start from the same set of initial
conditions. This allows to generate different `normalized' signals
such as {\mbox{$Q^{(b)}_{S}(\omega,t)=\sum_{j} S_{j}/\sum_{j}
s_{j}$}} or $Q^{(c)}_{S}(\omega,t)=\sum_{j} S_{j}/s_{j}/N_r$. These
normalized signals provide estimates that compensate to some degree
for the relatively small number of samples or for an~unknown mean
component but have the drawback that they cannot be applied in the
context of a physical experiment. Additionally, we divided these
signals by its means (time-averages) and removed the constant
component corresponding to the~zero-frequency response. Application
of the~Fourier-transform, $\widehat{Q}^{(b)}_{S} = \mathcal{F} \big(
Q^{(b)}_{S}/\overline{Q}^{(b)}_{S} -1\big)$ and
$\widehat{Q}^{(c)}_{S} = \mathcal{F}
\big(Q^{(c)}_{S}/\overline{Q}^{(c)}_{S}-1 \big)$, provides direct
access to amplitude and phase information.
Each of
these methods identified above yields the same general impression of
response maxima in time-modulated turbulence. Differences arise only
on a more detailed level of the processed data but these do not
obscure the interpretation of the main features of the response.
Therefore we only present results extracted from the normalized
signal $Q_{S}/\overline{Q}_{S} \equiv
Q^{(c)}_{S}/\overline{Q}^{(c)}_{S}$ in what follows,
unless explicitly
stated otherwise. The simulations were performed in the frequency
range $\pi/5 \leq \omega \leq 80\pi$ with time-modulated forcing at
an amplitude $A_F = 1/5$. Referring to Figure~\ref{fig:timemod0},
for each of~the~$N_{r}$ unmodulated initial conditions, $n_T=4$
periods of the modulated forcing were simulated, i.e., each sample
signal was computed for $n_{T}T$ time-units with modulation-period
$T=2\pi/ \omega$. Since an explicit time-stepping method was
adopted, the cases at low $\omega$ add particularly to the total
computational cost. The number of realizations required in the
ensemble was investigated separately. Results for several modulation
frequencies were compared at $N_{r}=10,~30$ and $50$; it was found
that $30$ independent samples provide adequate statistical
convergence for our purposes. We stored $N_{t}=40$ points per
modulation period and present results obtained by evaluating the
last two recorded periods, i.e., $2T \leq t \leq 4T$. Comparison
with results obtained by evaluating data on $0 \leq t \leq 4T$
yielded only minor differences.
Finally, the phase $\Phi_S (\omega )$ between the forcing and the response
can be computed from the Fourier-transformed data as well. At the
dominant frequency $f_A$ of the transformed signal
$\widehat{Q}_{S}(\omega,f) =
\mathcal{F}\big(Q_S(\omega,t))/\overline{Q}_S(\omega,t)-1\big)$, the
phase becomes $\Phi_S (\omega) = \arctan \big( {{\rm{Im}}
({\widehat{Q}_{S}}(\omega,{{f_A}}))} / {{\rm{Re}}
({\widehat{Q}_{S}}(\omega,{{f_A}}))}\big)$.

\begin{figure}[hbt]
\begin{center}
\includegraphics[width=0.45\textwidth]{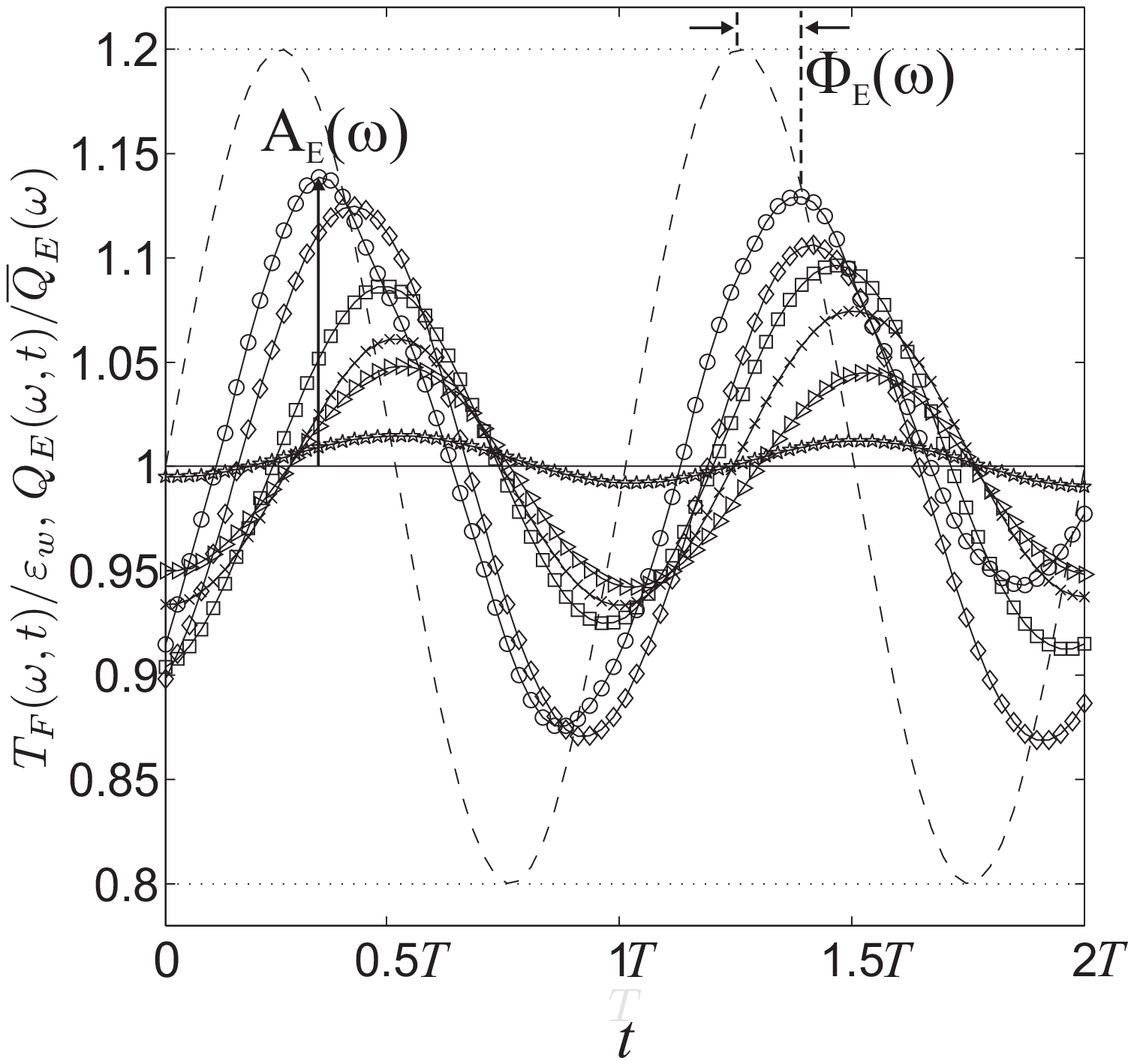}(a)
\includegraphics[width=0.45\textwidth]{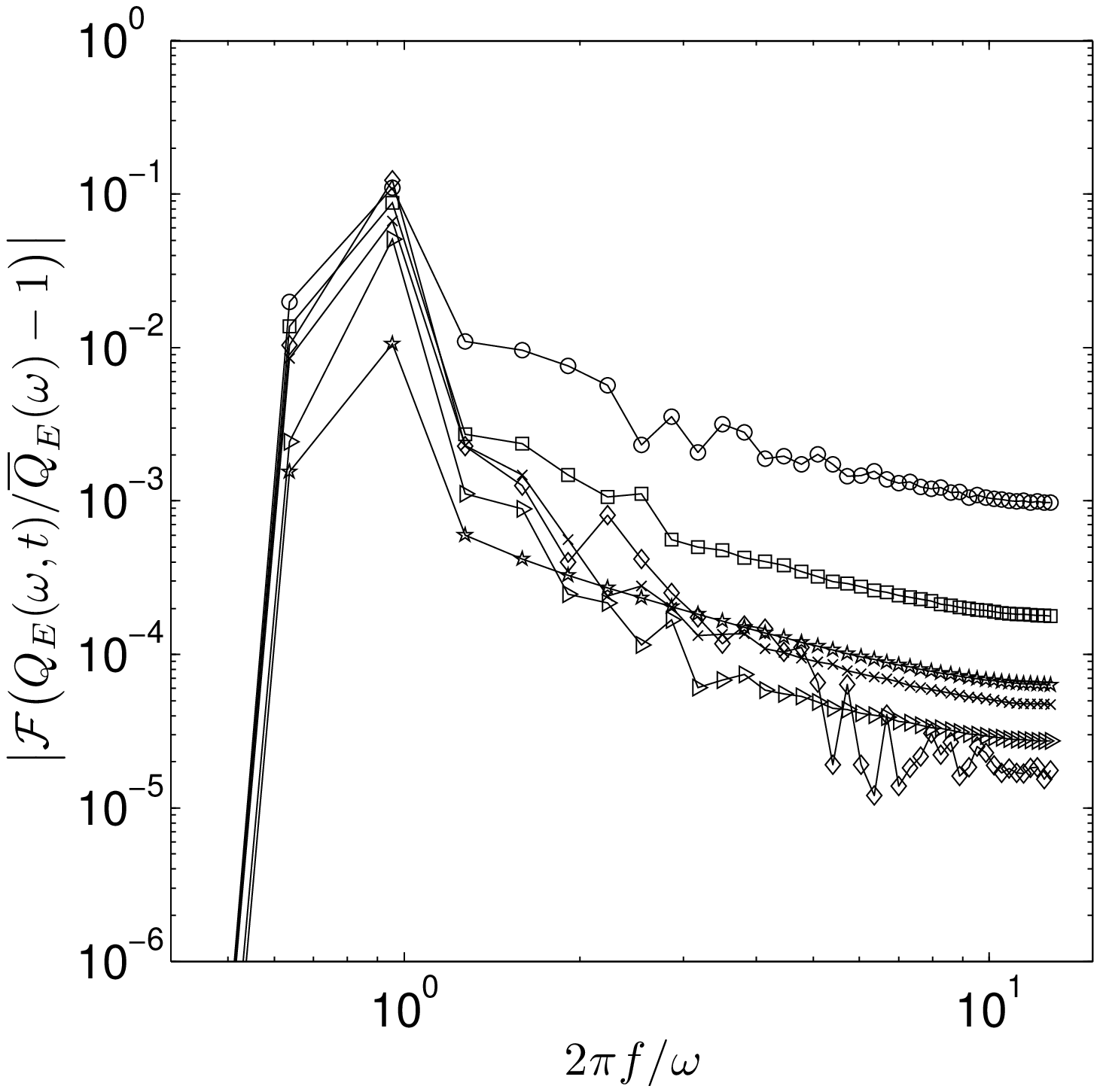}(b)
\end{center}
\caption{The response $Q_{E}(\omega,t)/\overline{Q}_E(\omega)$ for
the $R_{50}$ case recorded at different modulation frequencies $\omega$
is shown in (a) together with the modulation of the forcing
$T_F(\omega,t)/ \varepsilon_w$ (dashed). The corresponding power
spectra of the Fourier-transform as function of the transformed
frequency $f$ are collected in (b). Modulation frequencies
$\omega/(2\pi)=0.1, 0.2, 0.3, 0.4, 0.5, 2.0$ are included and
labeled by $\circ, \diamond, \square, \times, \triangleright$,
and~$\star$, respectively.}

\label{fig:xnorm}

\end{figure}

\section{Modulated turbulence}
In Figure~\ref{fig:xnorm}(a) the conditionally averaged signal
$Q_{E}(\omega,t)/\overline{Q}_{E}(\omega)$ based on total energy is
shown at a~number of~modulation frequencies. The conditionally
averaged response has a clear oscillatory behavior. The
Fourier-transform of the data from Figure~\ref{fig:xnorm}(a) is shown in
Figure~\ref{fig:xnorm}(b) and displays a dominant maximum
corresponding to the~forcing frequency $f_A = \omega/(2\pi)$. This
observation confirms that  the least-squares fitting
procedure adopted in \cite{hey03b} is justified.

\begin{figure}[hbt]
\begin{center}
\includegraphics[width=0.45\textwidth]{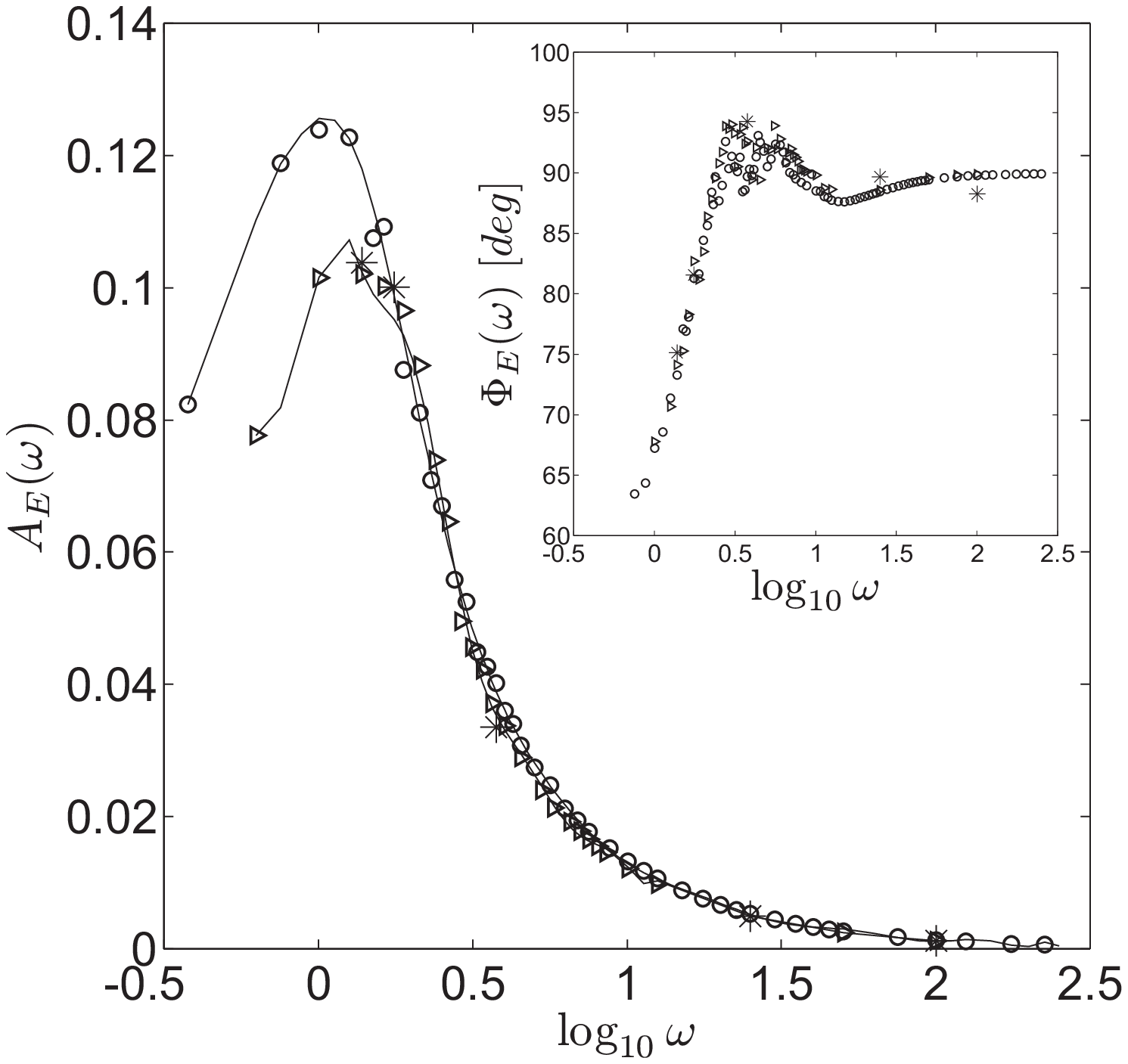}(a)
\includegraphics[width=0.45\textwidth]{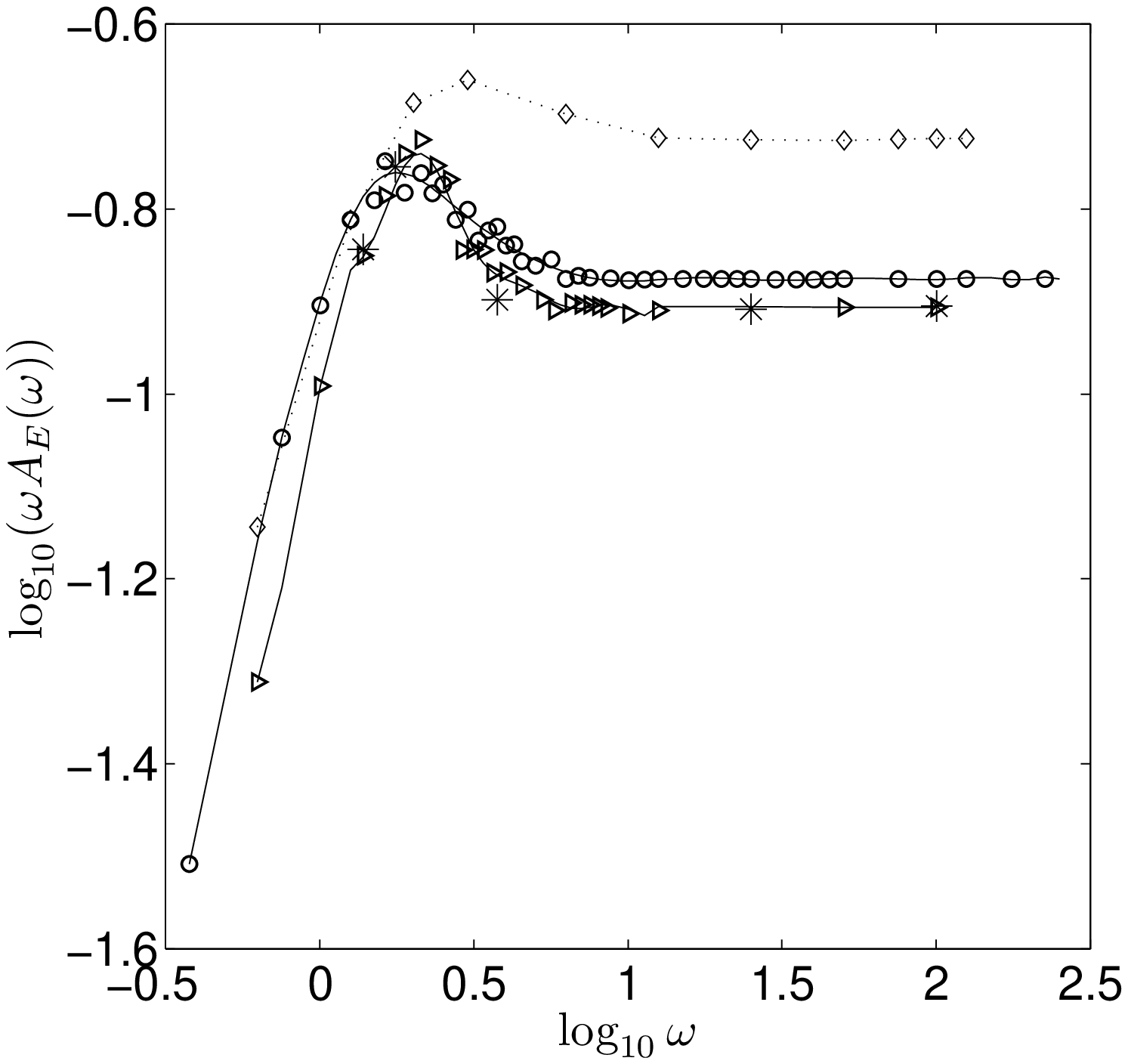}(b)
\end{center}
\caption{(a) Amplitude of the energy response $A_E(\omega)$ and (b)
compensated energy response $\omega A_E(\omega)$  obtained for the
$R_{50}$ (labeled~$\circ$) and the $R_{100}$
(labeled~$\triangleright$) cases. Verification at selected
frequencies for resolution $128^3$ and $R_{100}$ case
(labeled~$\star$). Results for forcing in two first shells ($k_F
\leq 5/2$) and the $R_{50}$ case (labeled~$\diamond$). The inset in
(a) shows the phase-shift $\Phi_E(\omega) $ between the~energy
response and the forcing modulation.} \label{fig:EandE}

\end{figure}

We now focus on the amplitude of the total energy response as
function of the modulation frequency
 $\omega$. The amplitude
${A_E}(\omega)$ computed as maximum of the Fourier-transformed
normalized signal for each modulation frequency is shown in
Figure~\ref{fig:EandE}(a). The maximum response appears at
$\omega_{max} \approx 1.5$, in accordance with the~expectation
\cite{hey03a,hey03b} that it should be close to the~inverse large
eddy turnover time. In addition, the~location of~the~maximum is not
very sensitive to $R_\lambda$, reflecting that the~response maximum
is mainly associated with the~large-scale features in the~flow.
At~high modulation frequencies $\omega > \omega_{max}$ the decay of
${A_E}$ is  proportional to $\omega^{-1}$, which becomes
particularly visible in the compensated response $\omega
{A_E}(\omega)$, Figure~\ref{fig:EandE}(b). At~very low modulation
frequencies $\omega < \omega_{max}$ a plateau in ${A_E}(\omega)$
must of course develop \cite{hey03a,hey03b}, as the turbulence then
completely follows the forcing. Our simulations do not achieve small
enough $\omega$ to observe a pronounced plateau.

The maximum of $\omega {A_E}(\omega)$ is about $35 \%$ higher as
compared to the value at high $\omega$. This is as expected lower
than predicted by the mean-field theory described in \cite{hey03a}
as the fluctuations slightly smear out the mean-field maximum, but
it is much more pronounced compared to results based on the GOY or
REWA simulations \cite{hey03b}. The reason is that, although
the~appearance of the~response maxima is a~large-scale effect,
the~correct resolution of the~small-scales is important for a~proper
quantitative representation of the effect, because the small scale
resolution affects the energy flux downscale. We also calculated
the~response curves for the~Taylor-Reynolds number; the results are
quite similar.

\begin{figure}[hbt]
\begin{center}
\includegraphics[width=0.45\textwidth]{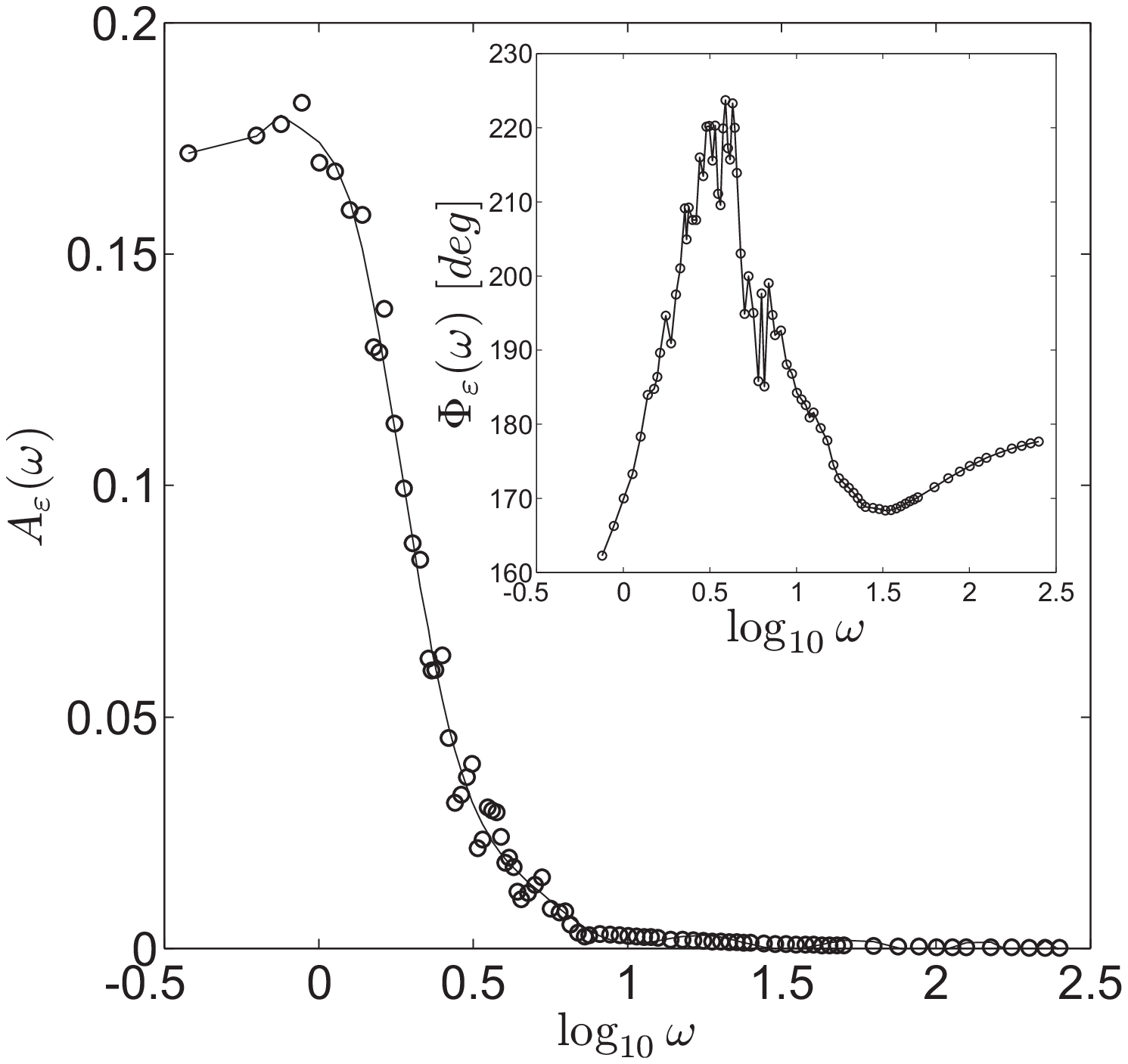}(a)
\includegraphics[width=0.45\textwidth]{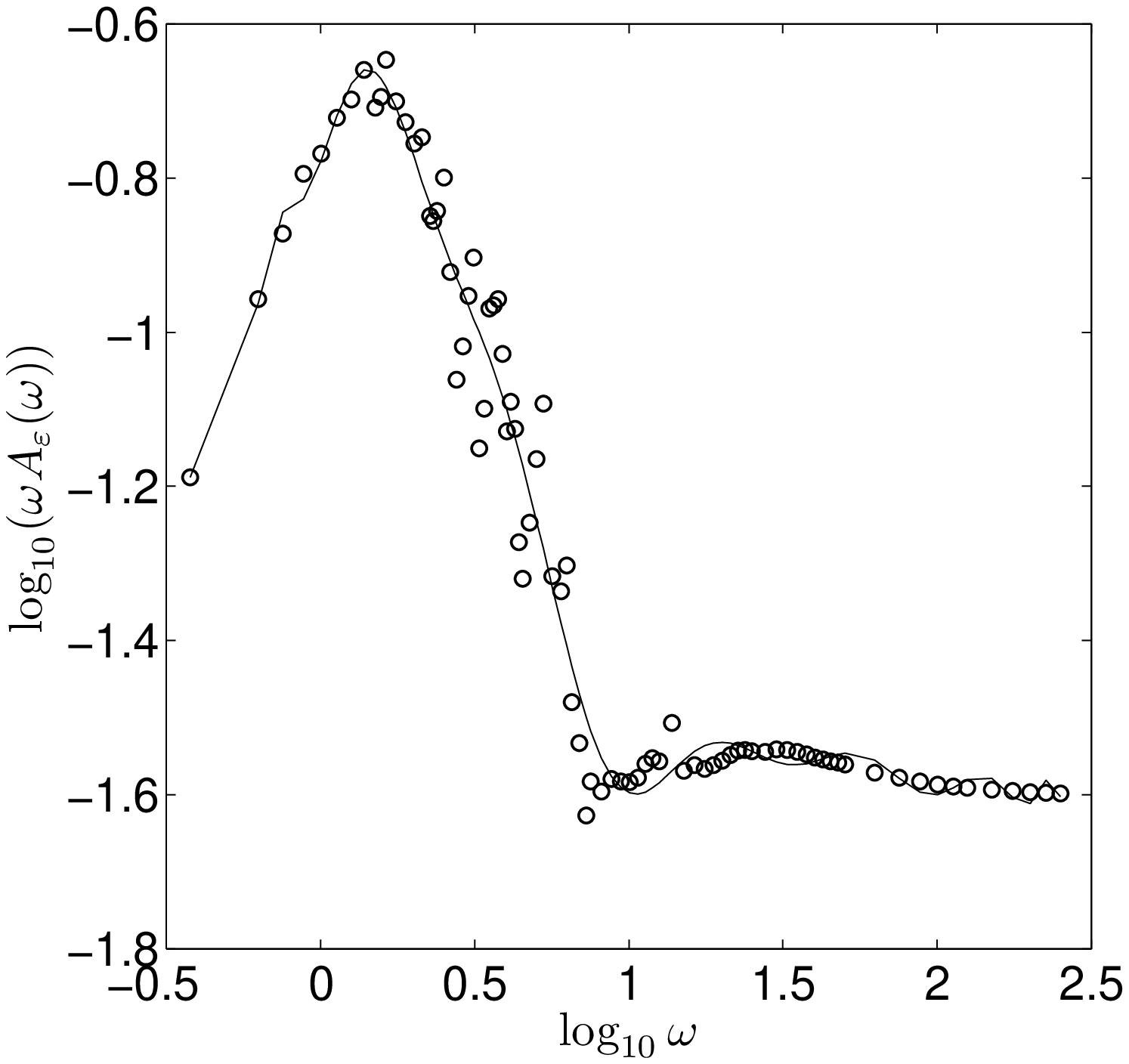}(b)
\end{center}
\caption{(a) Amplitude of the energy-dissipation-rate response
$A_\varepsilon(\omega)$
and (b) compensated energy-dissipation-rate response $\omega
A_\varepsilon(\omega)$  obtained for the $R_{50}$ case. The inset
in (a) shows the phase-shift
$\Phi_\varepsilon(\omega) $ between the energy-dissipation response
and the forcing modulation.}

\label{fig:EpsandEps}

\end{figure}

The phase-difference between the forcing modulation and
the~conditionally averaged total energy response is shown in
Figure~\ref{fig:EandE}(a) as inset. We observe a strong variation in
this phase-difference for modulation frequencies near the most
responsive modulation frequency. It~appears that the~maximum
response as shown in Figure~\ref{fig:EandE} occurs at a modulation
frequency where also the variation in the~phase-difference is
largest. A~strong phase shift was found similarly in windtunnel
experiments in which a~time-modulation is introduced via a~periodic
cycling of an upstream active grid. In these experiments the maximum
response was found to shift to higher frequencies in case the
characteristic length-scales of the forcing were reduced.

Can such a dependence on the type of~forcing also be observed in our
numerical simulations? To find  out we force a~higher wavenumber
band of modes ($k_F \leq 5/2$) instead of~restricting us entirely to
low wavenumber forcing. The~result is seen in
Figure~\ref{fig:EandE}(b) indicated by diamonds. Indeed, for this
type of~forcing the~response maximum is less pronounced. Further
quantitative connections with physical experiments are currently
being investigated.

The energy-dissipation-rate in the system is a~quantity that is
accessible to direct physical experimentation. In
Figure~\ref{fig:EpsandEps} we show the energy-dissipation-rate
amplitude $A_\varepsilon(\omega)$.
We notice that at high modulation frequency
$\omega$ the amplitude approaches zero, consistent with the
expectation that the~modulation of the forcing is not effective in
this range. More importantly, the~energy-dissipation-rate amplitude
displays a strong response maximum at the level of $85\%$ compared
to the amplitude of modulation. The~total mean energy-dissipation
$T^{-1} \int_0^{T} {\varepsilon (\omega ,t)dt}$
for each modulation frequency $\omega$ is almost constant. It
differs from the energy input rate $\varepsilon_w=0.15$ at the level
of $1 \%$ for most of the frequencies, reaching the maximum
difference of $5 \%$ for the lowest simulated frequency,
confirming good numerical convergence.

\section{Summary and Conclusions}
The direct numerical simulation of the response of turbulence to
time-modulated forcing confirms the existence of a response maximum.
The simulation findings are in general agreement with predictions
based on a mean-field theory \cite{hey03a}. The mean-field theory
predicts the decrease of the response amplitude proportional to
$\omega^{-1}$ as the modulation frequency is sufficiently large
which was observed in the simulations as well. The~response maxima
in the~total energy and the~Taylor-Reynolds number occur at the
forcing frequencies of the order of the inverse large eddy turnover
time scale.
The phase-difference between the
modulation of the forcing and the conditionally averaged response
displays a strong dependence on the modulation frequency as well.
The~modulation frequency at which the response maximum arises
depends only weakly on the~Reynolds number but shows a~dependence on
the scales included in the forcing as well as on the~flow-property
that is considered. In general, if the~particular quantity of
interest shows a~stronger dependence on the~smaller scales in
a~turbulent flow, then the response maximum arises at a~somewhat
higher frequency. These findings may be independently assessed in
physical experiments, e.g., conducted in wind tunnels combined with
the~use of active grids cycled in a~periodic sequence \cite{willem}.

\acknowledgments

Stimulating discussions with Willem van de Water (Eindhoven
University of Technology) are gratefully acknowledged. This work is
part of the research program `Turbulence and its role in energy
conversion processes' of the Foundation for Fundamental Research of
Matter (FOM), in the Netherlands, which is financially supported by
the Netherlands Organization for Scientific Research (NWO). The
authors wish to thank SARA Computing and Networking Services in
Amsterdam for providing the computational resources.


\begin{thebibliography}{10}

\bibitem{hey03a}
A. von~der Heydt, S. Grossmann, and D. Lohse, Phys. Rev. E {\bf 67},  046308
  (2003).

\bibitem{hey03b}
A. von~der Heydt, S. Grossmann, and D. Lohse, Phys. Rev. E {\bf 68},  066302
  (2003).

\bibitem{bif03}
L. Biferale, E. Calzavarini, F. Toschi, and R. Tripiccione,
Europhys. Lett. {\bf 64},  461  (2003).

\bibitem{bohr}
T. Bohr, M.~H. Jensen, G. Paladin, and A. Vulpiani, {\em Dynamical Systems
  Approach to Turbulence} (Cambridge University Press, Cambridge, 1998).

\bibitem{kad95}
L. Kadanoff, D. Lohse, J. Wang, and R. Benzi, Phys. Fluids {\bf 7},  617
  (1995).

\bibitem{egg91a}
J. Eggers and S. Grossmann, Phys. Fluids A {\bf 3},  1958  (1991).

\bibitem{gnlo92b}
S. Grossmann and D. Lohse, Z. Phys. B {\bf 89},  11  (1992).

\bibitem{gnlo94a}
S. Grossmann and D. Lohse, Phys. Fluids {\bf 6},  611  (1994).

\bibitem{gnlo94b}
S. Grossmann and D. Lohse, Phys. Rev. E {\bf 50},  2784  (1994).

\bibitem{cad03}
O. Cadot, J.~H. Titon, and D. Bonn, J. Fluid Mech. {\bf 485},  161  (2003).

\bibitem{gho95}
S. Ghosal, T.~S. Lund, P. Moin, and K. Akselvoll, J. Fluid Mech.
{\bf 286},  229  (1995).

\bibitem{esw88}
V. Eswaran and S.~B. Pope, Computers Fluids {\bf 16},  257  (1988).

\bibitem{pop00}
S.~B. Pope, {\em Turbulent Flow} (Cambridge University Press, Cambridge, 2000).

\bibitem{mey03}
J. Meyers, B.~J. Geurts, and M. Baelmans, Phys. Fluids {\bf 15},  2740  (2003).

\bibitem{geu03}
B.~J. Geurts, {\em Elements of direct and large-eddy simulation}
(R.T. Edwards, 2003).

\bibitem{willem}
W. van~der Water, 2005, private communication.

\end{thebibliography}
\end{document}